# Visualizing the Cu/Cu$_2$O Interface Transition in Nanoparticles with Environmental Scanning Transmission Electron Microscopy

Alec LaGrow,*,†,‡ Michael Ward,†,‡ David Lloyd,†,‡ Pratibha Gai,*,†,‡,∥ and Edward D. Boyes *,†,‡,§

†The York Nanocentre and Departments of ‡Physics, ∥Chemistry, and §Electronics, University of York, York YO10 5DD, U.K.

*S Supporting Information

**ABSTRACT:** Understanding the oxidation and reduction mechanisms of catalytically active transition metal nanoparticles is important to improve their application in a variety of chemical processes. In nanocatalysis the nanoparticles can undergo oxidation or reduction *in situ*, and thus the redox species are not what are observed before and after reactions. We have used the novel environmental scanning transmission electron microscope (ESTEM) with 0.1 nm resolution in systematic studies of complex dynamic oxidation and reduction mechanisms of copper nanoparticles. The oxidation of copper has previously been reported to be dependent on its crystallography and its interaction with the substrate. By following the dynamic oxidation process *in situ* in real time with high-angle annular dark-field imaging in the ESTEM, we 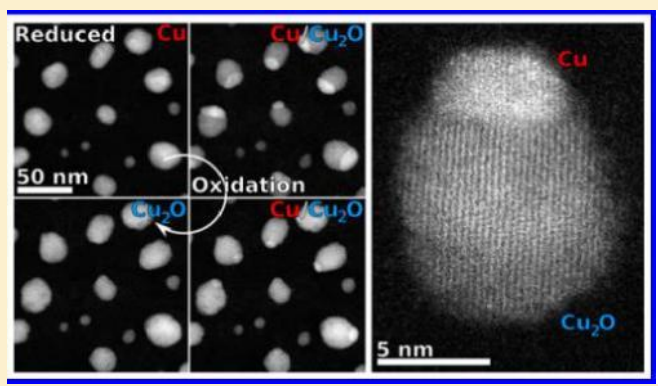 use conditions ideal to track the oxidation front as it progresses across a copper nanoparticle by following the changes in the atomic number (z) contrast with time. The oxidation occurs via the nucleation of the oxide phase (Cu$_2$O) from one area of the nanoparticle which then progresses unidirectionally across the particle, with the Cu-to-Cu$_2$O interface having a relationship of Cu{111}//Cu$_2$O{111}. The oxidation kinetics are related to the temperature and oxygen pressure. When the process is reversed in hydrogen, the reduction process is observed to be similar to the oxidation, with the same crystallographic relationship between the two phases. The dynamic observations provide unique insights into redox mechanisms which are important to understanding and controlling the oxidation and reduction of copper-based nanoparticles.

## ■ INTRODUCTION

A fundamental understanding of the oxidation and reduction mechanisms of metals is of critical importance to improving catalytic performance, corrosion control, and other industrial applications.[1] Copper is a particularly important material, as it is used in its metallic and oxidized forms for applications in electronics,[2,3] sensors,[4] and catalysis.[5] Copper is used commercially for methanol synthesis as part of the copper, alumina, and zinc oxide catalytic system.[6−8] For nanocatalysis, the catalytic processes are affected by the surface facets[8] and oxidation state,[5] and therefore it is important to understand how these change during the catalytic process.[9,10] In situ techniques allow us to gain insights into how a material behaves in reactive environments. Literature reports show that copper and its oxides undergo reduction,[11,12] oxidation,[6,13] and alloying over the course of catalytic reactions.[7] Understanding this behavior is of particular interest for preparing and stabilizing specific phases of copper.

In situ studies of Cu have focused on the oxidation of thin films.[1,14−19] The in situ transmission electron microscopy (TEM) studies have used conventional phase contrast TEM, mostly relying on dark-field imaging, electron diffraction, or high-resolution TEM to study the phase change from Cu to Cu$_2$O.[17−20] In thin films of Cu, the oxide was reported to nucleate on the films and then propagate along the Cu surface from a nucleation event until it reaches its maximum lateral size.[19] After this, the oxidation progresses to form Cu$_2$O islands on the Cu film.[17,18,20] The Cu$_2$O islands are seen to grow from each nucleation point and are crystallographically related to the underlying Cu,[1,20,21] with a further oxidation to CuO often seen ex situ with additional time[1,22,23] or temperature.[1,24] For Cu nanoparticles, the majority of reports show the eventual formation of hollow oxide nanoparticles via the Kirkendall effect.[22,25] The difference in the oxidation mechanisms between copper thin films and nanoparticles has not been fully explored, due in part to the difficulties in resolving the initial nucleation of oxide phases in nanoparticle species via regular *in situ* TEM techniques, as they generally need high-resolution imaging or information from scattering contrast.

In the present study, we introduce high-angle annular dark-field environmental scanning transmission electron microscopy (HAADF-ESTEM) to analyze the nucleation and propagation of the oxide through differences in atomic number (Z) contrast under reaction conditions. Dynamic *in situ* electron microscopy techniques have traditionally been carried out with TEM imaging in an environmental cell (ETEM),[26−29] but this limits the applicability of scanning transmission electron microscopy (STEM).[26,27] Recently, microelectromechanical systems (MEMS) have been incorporated into specialized holders for temperature control without drift,[30,31] as well as to support gas[32] and liquid environments.[33] In the gas cell holders, the gas is normally contained between two silicon nitride windows of ∼50 nm thickness, and they can reach pressures of 1 atm under static gas conditions or with low flow rates.[33] In the present work, we have used the novel environmental scanning transmission electron microscope (ESTEM),[34] which introduces gas directly into the column of a reconfigured microscope with a series of beamline differential pumping apertures.[35,36] The use of open differential apertures removes the need for gas-containing windows in the holder, and the ESTEM supports the full range of STEM functions, including high-sensitivity energy-dispersive X-ray spectroscopy (EDS) and single atom imaging. These advances allow for HAADF-STEM imaging of nanoparticles and even single atoms to be acquired under a subset of realistic conditions in gas and at temperature over the course of hours.

In this article, we report the oxidation of Cu and the reduction of $Cu_2O$ in situ using the environmental HAADF-STEM. In HAADF-ESTEM, the image intensity is approximately proportional to $Z^2$ and density, allowing the metal and its oxide to be readily distinguished, as seen in the $Cu/Cu_2O$ images in this paper and unlike in TEM. The oxidation and reduction processes are seen to be unidirectional, nucleating from a single area on the Cu or $Cu_2O$ nanoparticle surfaces and then spreading across the particle. The two phases are crystallographically related with the $Cu\{111\}//Cu_2O\{111\}$ relationship.

## EXPERIMENTAL SECTION

**Sample Preparation.** The Cu nanoparticles were prepared by sputtering 0.7 nm Cu at 80 mA[37] from a Cu target (99.99+% from Goodfellow Cambridge Ltd.) onto DENS Solutions Wildfire chips with ∼20 nm amorphous carbon films deposited with a JEOL JFC2300HR sputter coater. Before the experiments were started, the particles were reduced in the microscope at 500 °C with 2 Pa hydrogen (99.9995% from BOC UK) for 30 min to generate oxide-free metallic Cu.

**In Situ Analysis.** The analysis was carried out on a double-spherical aberration-corrected environmental (scanning) transmission electron microscope (AC-E(S)TEM, JEOL 2200FS) developed in-house by Boyes and Gai,[34−36] operating at 200 kV and equipped with a 100 $mm^2$ silicon drift detector (SDD) EDS detector from Thermo Scientific. The temperature was controlled with a MEMS heating stage from DENS Solutions. To avoid beam effects, including oxide reduction, calibration procedures[35] were employed to understand the maximum electron beam dose that could be applied to the system before damage occurred. The particles here were only exposed to the beam during data setup and acquisition, with 30 and 20.4 s exposures, respectively, for each image frame. The images were 1024×1024, with a pixel dwell time of 19.5 μs. For the in situ studies, the gas was introduced into the ESTEM with an inlet pressure 100 times higher than the specimen pressure (typically 200 Pa, giving a pressure of 2 Pa at the sample). High-purity oxygen gas (99.999% from BOC UK) was used for the oxidation and hydrogen gas (99.9995% from BOC UK) for the reduction.

**Simulations.** The interface model was made using VESTA[38] by combining Cu and $Cu_2O$ models into a single supercell. This model is a combination of two slabs, each with a (110) surface in the **c** axis. Both slabs were kept to the same thickness of approximately 10 nm:

39 unit cells for the Cu, and 33 unit cells of $Cu_2O$, with the extra atoms in the Cu slab trimmed to keep the thicknesses identical. The supercell was simulated using QSTEM[39] with the following parameters which best match the microscope: $C_s$ = 1 μm; convergence angle = 24 mrad; and HAADF detector range, 110−170 mrad. A total of 100×100 pixel array points were used, with a 200×200 probe array and a maximum scattering angle of 209 mrad. The focus was set to the top plane of the model. A total of 30 frozen phonon iterations were performed to take into account thermal diffuse scattering. The images presented use a source size of 1 Å and are oversampled by 10 times for presentation purposes.

## RESULTS AND DISCUSSION

The oxidation of copper was carried out at 300 °C with 2 Pa oxygen flowing onto the surface of the sample. An image was generally acquired every 10 min, with some noted exceptions, and the electron beam was blanked between exposures. In Figure 1, the oxidation front can be readily followed *in situ* by the change in Z-contrast that the oxidation causes, as the lattice expansion in three-dimensions accompanied by the lighter element oxygen causes a reduction in electron scattering intensity per unit area, and the oxidized area will be seen as having a substantially reduced HAADF image intensity. The darker area of each particle is the oxidized copper, and the brighter part is metallic copper.

The darker oxidized area nucleates from one edge of the nanoparticle and moves across each particle until, after 100 min, the particles are seen to be a single phase again. In each particle, only a single oxide nucleation event is seen, and the oxidation occurs from an initial nucleation point independent of nanoparticle size (Figure 1), as seen in all the particles observed (from 100 particles of ∼6 to ∼80 nm in size). The oxidation mechanism is indistinguishable on carbon and silicon nitride supports (Figure S1). From electron diffraction studies, the initial transition is observed to be from face-centered cubic (fcc) Cu to fcc $Cu_2O$ (Figure S2). The particles increased in size during the oxidation by ∼18% (from 100 particles), which, for a constant number of Cu atoms, is in agreement with the lattice expansion between Cu and $Cu_2O$. Cu and $Cu_2O$ have lattice parameters of 3.6148 and 4.2685 Å, respectively. EDS confirms the presence of oxygen in the dark side of the nanoparticles and not in the bright side (Figure S3).

The interface between the Cu and the $Cu_2O$ is observed to change with time, as measured from 10 particles *in situ* (Figure S4). After 20 min in 2 Pa oxygen at 300 °C, the interfaces are either flat or have an angle of 14 ± 4° into the Cu, with the $Cu_2O$ forming from a single side of the Cu, as noted by the white arrow in Figure 1. By 30 min, when approximately half of the particle has been oxidized, the Cu has an interface angle of 42 ± 10° into the $Cu_2O$ phase, as indicated by the arrow in Figure 1. On some nanoparticles, the $Cu/Cu_2O$ interface is terminated by separation (or de-wetting, Figure S4) of the two phases. This is shown in Figure 1, in which a wetting angle of 143° is observed after 30 min (indicated by the white arrow). At 60 min the angle is unchanged as the oxidation continues to move across the particle (Figure 1). At 80 min, just before full oxidation, the angle increases to 66° ± 13° (Figure 1). These changes would be difficult to detect in the TEM with phase contrast imaging, and the small differences in crystallography between Cu and $Cu_2O$. It should be noted that, due to STEM being a 2-D projection of a 3-D space, it is difficult to tell the exact angular representation with time.

At 500 °C, the particles were seen to undergo the same oxidation mechanism at an increased oxidation rate (Figure 2).

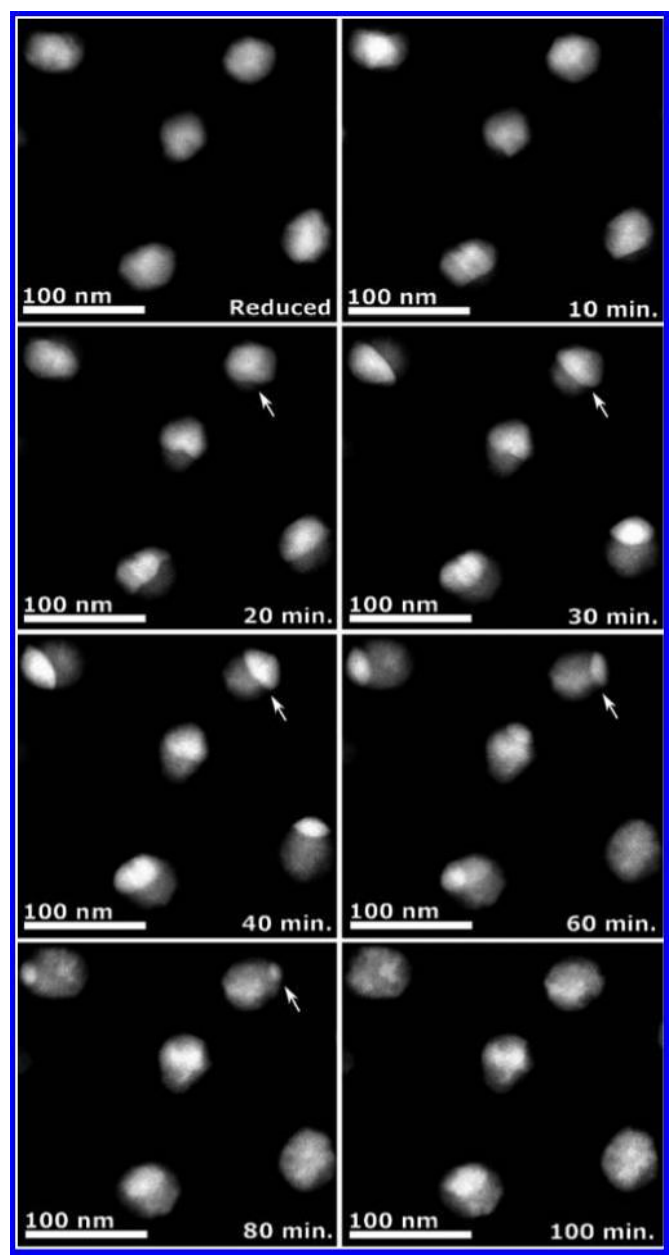

Figure 1. In situ oxidation of Cu on carbon support over the course of 100 min at 300 °C in 2 Pa oxygen. The white arrow points to a nanoparticle that shows the shrinking Cu area, formation of Cu$_2$O, and the change in Cu-to-Cu$_2$O interface angles.

The nanoparticle oxidized in Figure 2 is shown to have the oxidation front move perpendicular to the viewing direction, and thus the angle can be accurately analyzed. The interfaces were seen to move across the particles as the oxide formed, until the particles were a single phase of Cu$_2$O (blue arrow in Figure 2). The oxide that was formed showed a roughening in surface structure as the particles lost their distinctive faceting (Figure 2, 33−40 min). The interface became curved at 35 min, with an angle of 4°, and at 40 min the angle on one side of the Cu changed to 83° (indicated by the white arrow) as the Cu$_2$O oxidized around the Cu (Figure 2). It should also be noted that, at 500 °C, the images were taken every 2.5 min to follow the reaction and that, with the increased dose during acquisition, beam effects become evident, with sintering and moving of the nanoparticle during the imaging process. This is particularly noticeable with the migration of the smaller particle into the imaging frame that appears after 30 min and the rotation of the imaged particle.

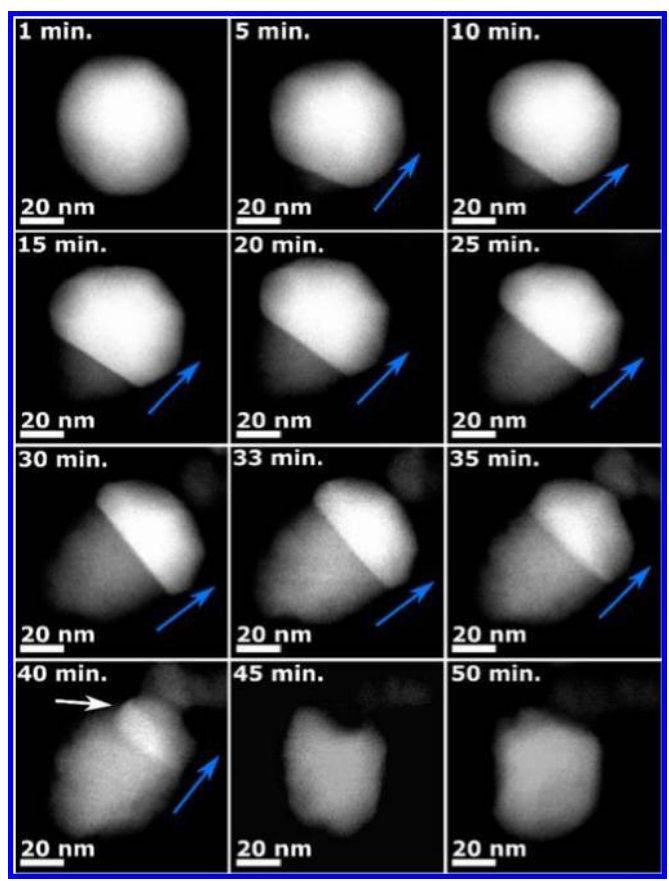

Figure 2. In situ oxidation of Cu carried out at 500 °C in 2 Pa oxygen. The blue arrow indicates the direction in which the interface moves.

A similar oxidation process also occurred when the reduced copper nanoparticles were placed *ex situ* in a desiccator for a week at room temperature. The particles showed the same two phases as seen *in situ*; however, they also showed a degree of hollowing between the Cu and Cu$_2$O (Figure S5). This oxidation mechanism is also similar to literature reports of the synthesis of Cu/Cu$_2$O semishell nanoparticles.[40]

The oxidation procedure was seen to be pressure dependent. At higher pressures (up to 10 Pa pressure at the sample), the reaction was seen to occur much quicker, within 5 min. The nucleation and oxidation mechanism was, however, unchanged with pressure, with the oxidation occurring from one area on the nanoparticle and the oxide front moving unidirectionally across the nanoparticle (Figure S6). In this pressure range, the reaction kinetics increased with temperature and pressure, and the process had an exponential relationship between the gas pressure (a linear relationship with reciprocal pressure) and the oxidation rate (Figure S6). The higher pressure reaction also showed that, at longer oxidation times, the particles hollowed and voids formed (Figure S7). The hollowing was similar to previous reports of the oxidation of Cu.[22,41]

The fully oxidized nanoparticles from Figure 1 were then studied during their reduction by replacing the oxygen atmosphere with 2 Pa of hydrogen. The reduction occurred in a similar way to the oxidation, with the reduced phase nucleating at the surface of the oxidized phase and moving across the nanoparticle (Figure 3). The Cu formed as a protruding island off the nanoparticle, with a de-wetting angle (Figure S4) of 144° from the Cu2O surface and an interface length of ∼15 nm, measured from the particle indicated with the white arrow in Figure 3. As the temperature was increased, the size of the interface between the two phases became larger and the protrusion lessened, with the interfacial area expanding to ∼25 nm as the temperature was increased to 400 °C (Figure 3) and the reduction occurred. As the temperature reached 500 °C, the nanoparticle had a larger interface as the reduction went to completion and the de-wetting of the Cu on the Cu2O disappeared.

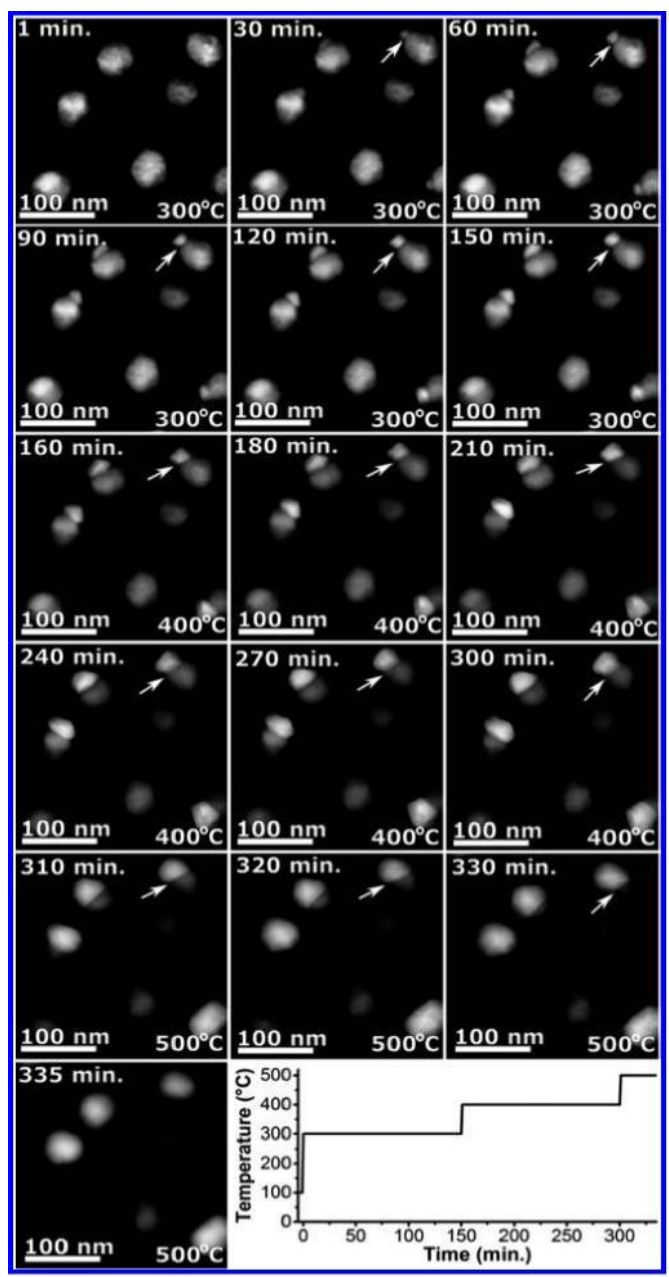

Figure 3. In situ reduction of Cu2O in 2 Pa hydrogen over the course of 335 min. In the lower right-hand corner, the temperature profile of the reaction is shown.

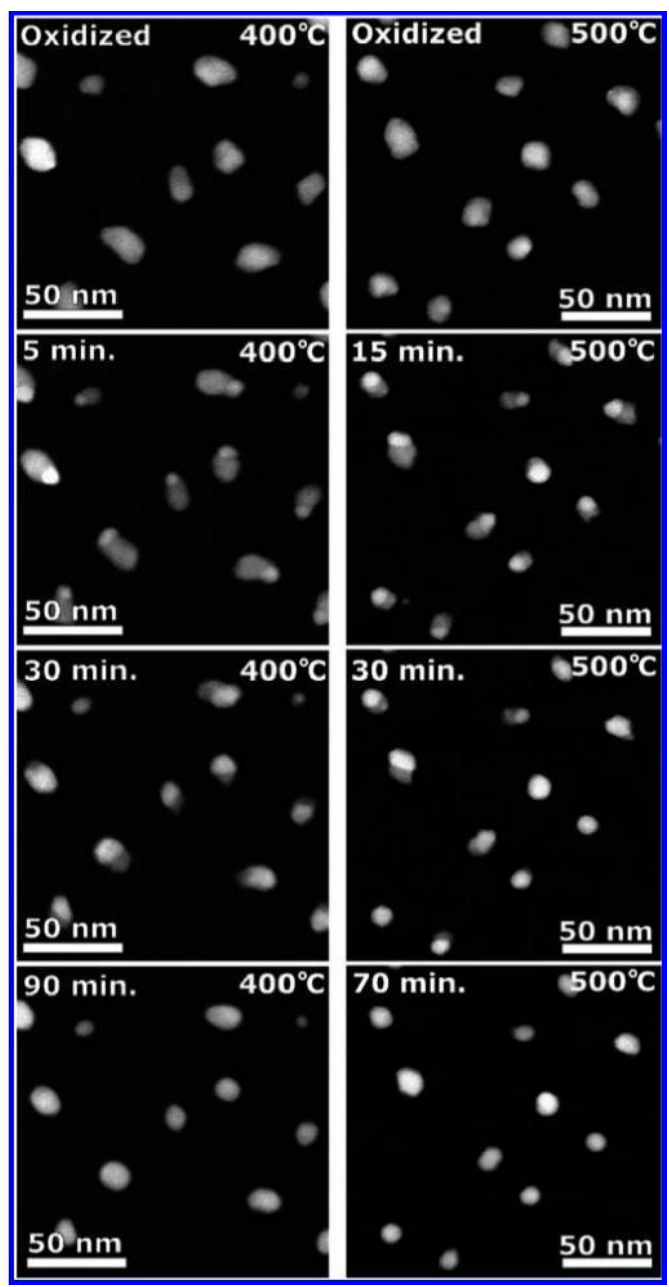

Figure 4. In situ reduction of Cu2O in 2 Pa hydrogen at 400 °C in the left column and at 500 °C in the right column

The reduction experiments were also carried out at 400 and 500 °C (Figures 4, S8, and S9). At 400 and 500 °C, the nanoparticles had a reduction mechanism with very similar behavior to the oxidation, and there was a single interface between the two phases. The appearance of de-wetting between the Cu areas to the Cu2O of 147 ± 20° at 400 °C and 164 ± 12° at 500 °C, was measured in the early stages of reduction from 10 particles in each experiment. The reduction was also seen to occur in a similar fashion from hollowed nanoparticles, with the Cu nucleating on one side and the reduction moving across the particle as the Cu phase formed (Figure S10). Electron diffraction studies showed that the process was the reduction of Cu2O to Cu (Figure S11). From these results, it is shown that the reduction rate is dependent on temperature. It should be noted that there is no memory effect, in that the nucleation sites during oxidation and reduction are not related to each other in subsequent cycles (Figure S12).

To probe further the interface between the Cu and Cu2O, high-resolution images from different stages of the reaction were analyzed. The high-resolution dynamic HAADF-ESTEM image in Figure 5A shows lattice spacings of the nanoparticle corresponding to the (111) spacing of Cu (2.1 Å) and the (111) spacing of Cu$_2$O (2.5 Å). The two spacings are seen to be directly related to each other with parallel epitaxy between the Cu and Cu$_2$O fcc crystal structures. The relationship between the Cu/Cu$_2$O has a {111}//{111} crystallographic relationship across all particles, even though there is an 18% lattice expansion between the two phases (Figure 5A). The FFT of the image indicated that it had a 2.5° rotation between the Cu and Cu$_2$O planes (Figure 5B), which is consistent with the average of 10 Cu/Cu$_2$O interfaces showing rotations of 2.5 ± 1.5° between the two planes. The Cu{111}//Cu$_2$O{111} relationship was seen to be constant across all oxidized (Figure S13) and reduced (Figure S14) particles. Line profiles were carried out on either side of the interface to understand the lattice matching relationship between the Cu and Cu$_2$O. Figure 5C shows a particle with an interface length of ∼6 nm, with a relationship between the Cu and Cu$_2$O of 7 (111) planes of Cu matched to 6 (111) planes of Cu$_2$O, indicating that the interface had a 7✕6 lattice matching relationship (Figure 5D). For the particle during reduction with an interface of ∼14 nm (Figure 5E), a 6✕5 lattice matching relationship was seen to be dominant, with units of the 7✕6 interface near the center of the particle (Figure 5G). In additional particles during oxidation, the 7✕6 behavior was seen in interfaces of ∼6 nm, and in reduction with a larger interface the 6✕5 relationship was seen (Figure S15).

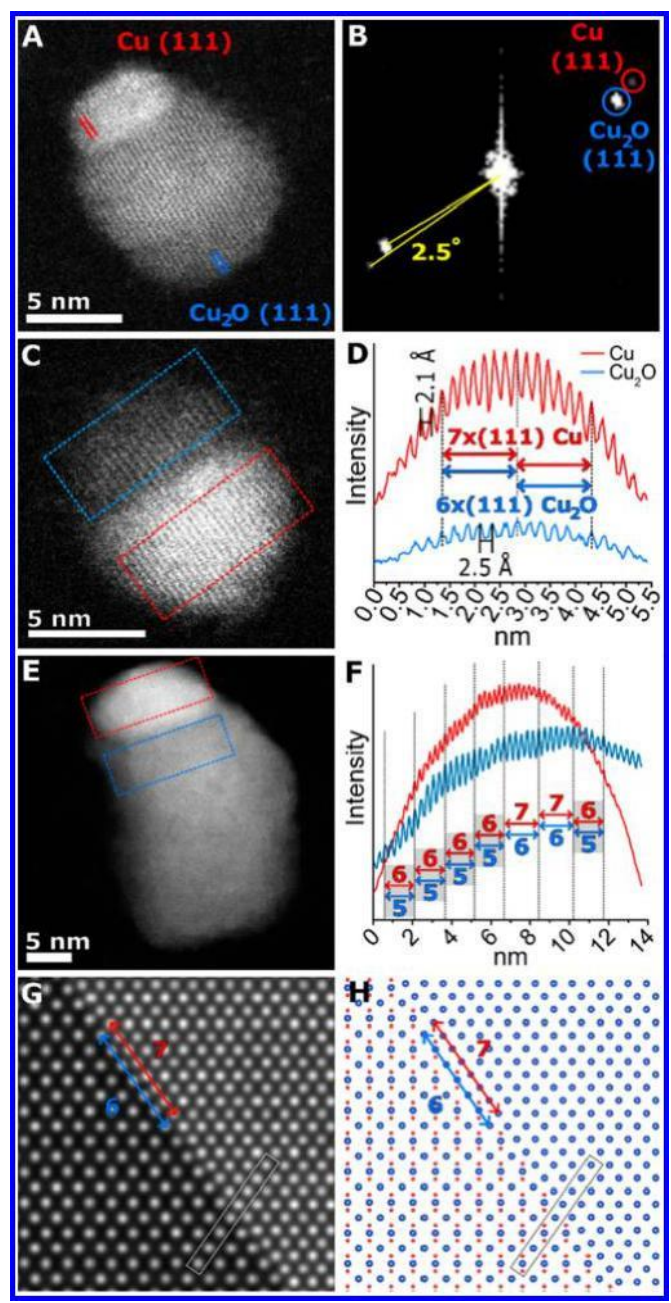

Figure 5. (A) Dynamic HAADF-ESTEM image of a Cu/Cu$_2$O nanoparticle, with (B) the FFT showing the 2.5° rotation between the (111) reflections of Cu and Cu$_2$O. HAADF-STEM images of (C) a Cu/Cu$_2$O nanoparticle and (E) a Cu/Cu$_2$O nanoparticle during reduction. (D,F) Line profiles from the red and blue boxes in panels C and E, with annotations indicating the lattice matching of the Cu/Cu$_2$O, with a gray background behind the 6✕5 relationship. (G) QSTEM simulation of the Cu to Cu$_2$O interface model shown in panel H, with the 7✕6 interface highlighted in both, and the attachment area highlighted with a white box in panel G and a black box in panel H.

A simulation of the interface was carried out in QSTEM (Figure 5G) by matching the Cu and Cu$_2$O model (Figure 5H) surfaces, both viewed down the [110] direction. The model shows the 7✕6 Cu/Cu$_2$O lattice relationship (Figure 5G, annotated with arrows and in the white box), with the aligned copper atoms at the end of each row (Figure 5H). The oxygen atoms were seen to be aligned between the layers in the interstitial tetrahedral sites (Figure 5H). The HAADF-STEM images seen in Figure 5A,C,E would be tilted away from the ideal zone axis, as only the (111) planes are visible. This is consistent with models carried out at increased tilt angles (Figure S16). These results illustrate the sensitivity of image contrast, especially at the inter-phase interface, to the small angles of specimen tilt misalignment of nanoparticles in the microscope. As a consequence, the images of even sharp interfaces will inevitably appear in images to be more diffuse than is actually the case.

The difference between atomic structures of the Cu and Cu$_2$O is that the Cu$_2$O has oxygen atoms in the interstitial tetrahedral sites (Figure S17). The oxygen atoms essentially are positioned between the {111} planes of copper. This relationship between the atoms indicates why the relationship between the Cu and Cu$_2$O is observed to be Cu{111}//Cu$_2$O{111}, as the oxygen could pack into the interstitial tetrahedral sites, causing a lattice expansion, while retaining the underlying fcc crystal structure with the inclusion of oxygen. In the [110] zone axis, where the (111) planes can be resolved, the oxygen is seen to be off-axis to the copper planes, and thus this would not contribute directly to the Z contrast of the atomic columns (Figure S17), leading to the reduction in contrast with oxidation.

Cu{111}//Cu$_2$O{111} is a common relationship between the two phases in the literature.[1,42] Previous studies have shown that a 7✕6 misfit provides the minimum coincidence misfit of 1.22%.[21,43] The 7✕6 relationship represents 3 times the O−O distance in Cu$_2$O (111) to 7 times the Cu−Cu distance.[44] The slight orientation difference between the two phases is similar, but smaller than the ∼7% change in orientation reported by Milne and Howie in copper oxide islands grown on thin films.[17] The 7✕6 relationship and the rotation of ∼2.5° between the two phases would reduce the strain between the mismatched phases causing the Cu/Cu$_2$O particles to be stable during oxidation and allowing the stable Cu{111}//Cu$_2$O{111} relationship to exist. The 6✕5 relationship has also been widely reported[18,45,46] and has a misfit of 2.02% in the opposite direction. The combination of the two lattice matching types would have the minimum possible strain on the interface. In this study on nanoparticles, the dominant relationship in the smaller nanoparticles (∼6 nm) was the 7×6 mismatch, while a mixed phase was observed of 7×6 and 6×5 in the larger particles during reduction.

In previous studies with thin films, it is reported that, when the Cu surface is exposed to oxygen, a rearrangement occurs into a Cu−O surface.[1,15,47] The oxygen then dissociates and can diffuse along the surface, evaporate, or form a nucleus.[1,15] Once a nucleus is formed, oxygen preferentially diffuses to the nucleus, and the oxygen concentration around the nucleus creates a zone of capture.[1,15] The oxidation then progresses with either the copper diffusing out toward the oxygen layer, or the oxygen diffusing into the structure.[15] Previous reports showed that the oxygen creates $Cu^+$ cations that readily diffuse instead of the oxygen diffusing.[48,49] The oxidation then occurs by a layer-by-layer approach.[14,49] In this study, we have shown the formation of a single oxide nucleus, in the observed particles, which grows as the particles are oxidized. The oxidation occurs at the interface, indicating that the copper or oxygen diffuses preferentially instead of creating new oxide nuclei on the surface. This is in agreement with thin-film studies, where the oxidation occurs preferentially at nucleation sites instead of creating new oxide nuclei.[1,15]

For the reduction of $Cu_2O$, the reduction occurs in a similar manner to the oxidation. Literature reports that the reduction occurs via hydrogen embedding into the oxide lattice, weakening the Cu−O bond, which would then form Cu−OH bonds and further catalyze the removal of oxygen.[50] Previous reports show that the reduction of copper oxide films occurs at the surface and progressively moves into the thin film.[51] In situ studies of copper oxide powders have indicated that the reduction occurs by forming islands of Cu and the reaction progresses at the interface between the copper and its oxide.[50] Our results from nanoparticles are in agreement with the earlier thin-film studies,[50,51] and we show that, once an initial nucleus is formed, the reduction occurs preferentially at the interface between the $Cu_2O$ and Cu.

In the case of copper nanoparticles, it is shown that the oxidation and reduction occurs from an initial nucleation event of either the metal or the oxide, and the reaction kinetics is then interface mediated. The two phases are crystallographically related in the oxidation and the reduction, and the strain of the lattice mismatch is minimized by adopting a 7×6, 6×5, or mixed 7×6 and 6×5 relationship between the two phases.

## CONCLUSIONS

In conclusion, the introduction of atomic-resolution HAADF-ESTEM allowed the initial stages of the processes to be studied during the oxidation of Cu nanoparticles and reduction of $Cu_2O$ nanoparticles. The oxidation of Cu to $Cu_2O$ occurred by the nucleation of $Cu_2O$ from one point on each Cu nanoparticle, with the oxidation front moving unidirectionally across the nanoparticle with time. The same oxidation mechanism could be seen from 300 to 500 °C and at 2, 5, and 10 Pa pressure, and it was consistent with ex situ measurements. The reaction was reversible in hydrogen and occurred in a similar way, although the wetting angle between the copper and its oxide was much lower, creating a protrusion off the particle to minimize the interface size, with island formation occurring at 300 °C in hydrogen. The wetting angle was increased at higher temperatures (400 and 500 °C) during the reduction, minimizing the extent of protrusion of the copper phase. The in situ observations of the faster oxidation of copper metal and the slower reduction of the oxide show that the two processes are asymmetrical. The interface between the Cu and $Cu_2O$ was observed to have a Cu{111}//$Cu_2O${111} relationship with either the 7×6 or 6×5 lattice matching relationships, or a mixture of the two. These results demonstrate the value of using atomic-resolution in situ HAADF-ESTEM to follow the oxidation and reduction fronts by the changes in Z contrast during reaction in gas environments and at high temperatures.

## ■ ASSOCIATED CONTENT

\* Supporting Information

Electron diffraction, additional experiments, and additional high-resolution HAADF-STEM images, including Figures S1−S17 (PDF)

## ■ AUTHOR INFORMATION

Corresponding Authors
\*alec.lagrow@york.ac.uk
\*pratibha.gai@york.ac.uk
\*ed.boyes@york.ac.uk
ORCID
Edward D. Boyes: 0000-0001-8456-1208
Notes
The authors declare no competing financial interest.

## ■ ACKNOWLEDGMENTS

P.G. and E.B. thank the EPSRC (UK) for a critical mass Grant EP/J0118058/1, for postdoctoral research assistantships (PDRAs) for A.L. and M.W. from the grant, and a Ph.D. studentship for D.L. We thank Ian Wright for expert technical assistance.

## ■ REFERENCES

(1) Gattinoni, C.; Michaelides, A. Surf. Sci. Rep. 2015, 70, 424.
(2) Dubal, D. P.; Dhawale, D. S.; Salunkhe, R. R.; Jamdade, V. S.; Lokhande, C. D. j Alloys Compd. 2010, 492, 26.
(3) Brittman, S.; Yoo, Y.; Dasgupta, N. P.; Kim, S.-i.; Kim, B.; Yang, P. Nano Lett. 2014, 14, 4665.
(4) Chen, W.; Chen, J.; Feng, Y.-B.; Hong, L.; Chen, Q.-Y.; Wu, L.-F.; Lin, X.-H.; Xia, X.-H. Analyst 2012, 137, 1706.
(5) Gawande, M. B.; Goswami, A.; Felpin, F.-X.; Asefa, T.; Huang, X.; Silva, R.; Zou, X.; Zboril, R.; Varma, R. S. Chem. Rev. 2016, 116, 3722.
(6) Gai, P. L.; Smith, B. C.; Owen, G. Nature 1990, 348, 430.
(7) Wagner, J. B.; Hansen, P. L.; Molenbroek, A. M.; Topsøe, H.; Clausen, B. S.; Helveg, S. j Phys. Chem. B 2003, 107, 7753.
(8) Behrens, M.; Studt, F.; Kasatkin, I.; Kühl, S.; Hävecker, M.; Abild-Pedersen, F.; Zander, S.; Girgsdies, F.; Kurr, P.; Kniep, B.-L.; Tovar, M.; Fischer, R. W.; Nørskov, J. K.; Schlögl, R. Science 2012, 336, 893.
(9) Hansen, P. L.; Wagner, J. B.; Helveg, S.; Rostrup-Nielsen, J. R.; Clausen, B. S.; Topsøe, H. Science 2002, 295, 2053.
(10) Uchiyama, T.; Yoshida, H.; Kuwauchi, Y.; Ichikawa, S.; Shimada, S.; Haruta, M.; Takeda, S. Angew. Chem., Int. Ed. 2011, 50, 10157.
(11) Sinatra, L.; LaGrow, A. P.; Peng, W.; Kirmani, A. R.; Amassian, A.; Idriss, H.; Bakr, O. M. j Catal. 2015, 322, 109.
(12) Peppley, B. A.; Amphlett, J. C.; Kearns, L. M.; Mann, R. F. Appl. Catal., A 1999, 179, 31.
(13) Deng, Y.; Handoko, A. D.; Du, Y.; Xi, S.; Yeo, B. S. ACS Catal. 2016, 6, 2473.

(14) Yang, J. C.; Kolasa, B.; Gibson, J. M.; Yeadon, M. *Appl. Phys. Lett.* 1998, 73, 2841.
(15) Zhou, G.; Yang, J. C. *Surf. Sci.* 2003, 531, 359.
(16) Heinemann, K.; Rao, D. B.; Douglass, D. L. *Oxid. Met.* 1975, 9, 379.
(17) Milne, R. H.; Howie, A. *Philos. Mag. A* 1984, 49, 665.
(18) Zhou, G.; Luo, L.; Li, L.; Ciston, J.; Stach, E. A.; Saidi, W. A.; Yang, J. C. *Chem. Commun.* 2013, 49, 10862.
(19) Li, L.; Luo, L.; Ciston, J.; Saidi, W. A.; Stach, E. A.; Yang, J. C.; Zhou, G. *Phys. Rev. Lett.* 2014, 113, 136104.
(20) Luo, L.; Kang, Y.; Yang, J. C.; Zhou, G. *Surf. Sci.* 2012, 606, 1790.
(21) Zhou, G. W. *Acta Mater.* 2009, 57, 4432.
(22) Nakamura, R.; Tokozakura, D.; Lee, J. G.; Mori, H.; Nakajima, H. *Acta Mater.* 2008, 56, 5276.
(23) Platzman, I.; Brener, R.; Haick, H.; Tannenbaum, R. *J. Phys. Chem. C* 2008, 112, 1101.
(24) O'Reilly, M.; Jiang, X.; Beechinor, J. T.; Lynch, S.; NíDheasuna, C.; Patterson, J. C.; Crean, G. M. *Appl. Surf. Sci.* 1995, 91, 152.
(25) Susman, M. D.; Vaskevich, A.; Rubinstein, I. *J. Phys. Chem. C* 2016, 120, 16140.
(26) Boyes, E. D.; Gai, P. L. *Ultramicroscopy* 1997, 67, 219.
(27) Jinschek, J. R. *Chem. Commun.* 2014, 50, 2696.
(28) Sharma, R.; Weiss, K. *Microsc. Res. Tech.* 1998, 42, 270.
(29) Gai, P. L. *Adv. Mater.* 1998, 10, 1259.
(30) van Huis, M. A.; Kunneman, L. T.; Overgaag, K.; Xu, Q.; Pandraud, G.; Zandbergen, H. W.; Vanmaekelbergh, D. *Nano Lett.* 2008, 8, 3959.
(31) Damiano, J.; Nackashi, D.; Mick, S. *Microsc. Microanal.* 2008, 14, 1332.
(32) Allard, L. F.; Bigelow, W. C.; Bradley, S. A.; Liu, J. *Microsc. Today* 2009, 17, 50.
(33) Jonge, N. d.; Peckys, D.; Veith, G.; Mick, S.; Pennycook, S.; Joy, D. *Microsc. Microanal.* 2007, 13, 242.
(34) Boyes, E. D.; Ward, M. R.; Lari, L.; Gai, P. L. *Ann. Phys. (Berlin)* 2013, 525, 423.
(35) Boyes, E. D.; Gai, P. L. *C. R. Phys.* 2014, 15, 200.
(36) Gai, P. L.; Boyes, E. D. *Microsc. Res. Tech.* 2009, 72, 153.
(37) Martin, T. E.; Gai, P. L.; Boyes, E. D. *ChemCatChem* 2015, 7, 3705.
(38) Momma, K.; Izumi, F. *J. Appl. Crystallogr.* 2011, 44, 1272.
(39) Koch, C. T., Arizona State University, 2002.
(40) Giannousi, K.; Sarafidis, G.; Mourdikoudis, S.; Pantazaki, A.; Dendrinou-Samara, C. *Inorg. Chem.* 2014, 53, 9657.
(41) Susman, M. D.; Vaskevich, A.; Rubinstein, I. *J. Phys. Chem. C* 2016, 120, 16140.
(42) Ronnquist, A.; Fischmeister, H. *J. Inst. Metals* 1960, 89, 65.
(43) Chu, Y. S.; Robinson, I. K.; Gewirth, A. A. *J. Chem. Phys.* 1999, 110, 5952.
(44) Chen, C.-H.; Yamaguchi, T.; Sugawara, K.-i.; Koga, K. *J. Phys. Chem. B* 2005, 109, 20669.
(45) Perez Leon, C.; Surgers, C.; Löhneysen, H. v. *Phys. Rev. B: Condens. Matter Mater. Phys.* 2012, 85, 035434.
(46) Matsumoto, T.; Bennett, R. A.; Stone, P.; Yamada, T.; Domen, K.; Bowker, M. *Surf. Sci.* 2001, 471, 225.
(47) Tobin, J. G.; Klebanoff, L. E.; Rosenblatt, D. H.; Davis, R. F.; Umbach, E.; Baca, A. G.; Shirley, D. A.; Huang, Y.; Kang, W. M.; Tong, S. Y. *Phys. Rev. B: Condens. Matter Mater. Phys.* 1982, 26, 7076.
(48) Bardeen, J.; Brattain, W. H.; Shockley, W. *J. Chem. Phys.* 1946, 14, 714.
(49) Cabrera, N.; Mott, N. F. *Rep. Prog. Phys.* 1949, 12, 163.
(50) Kim, J. Y.; Rodriguez, J. A.; Hanson, J. C.; Frenkel, A. I.; Lee, P. L. *J. Am. Chem. Soc.* 2003, 125, 10684.
(51) Li, J.; Vizkelethy, G.; Revesz, P.; Mayer, J. W.; Tu, K. N. *J. Appl. Phys.* 1991, 69, 1020.

Supporting Information for:

# Visualizing the Cu/Cu$_2$O Interface Transition in Nanoparticles with Environmental Scanning Transmission Electron Microscopy

Alec P. LaGrow,*,[†,‡] Michael R. Ward,[†,‡] David C. Lloyd,[†,‡] Pratibha L. Gai,*,[†,‡] and Edward D. Boyes*,[†,‡,§]

[†]The York Nanocentre, [‡]Department of Physics, Chemistry, and [§]Electronics, University of York, York YO10 5DD, U.K.

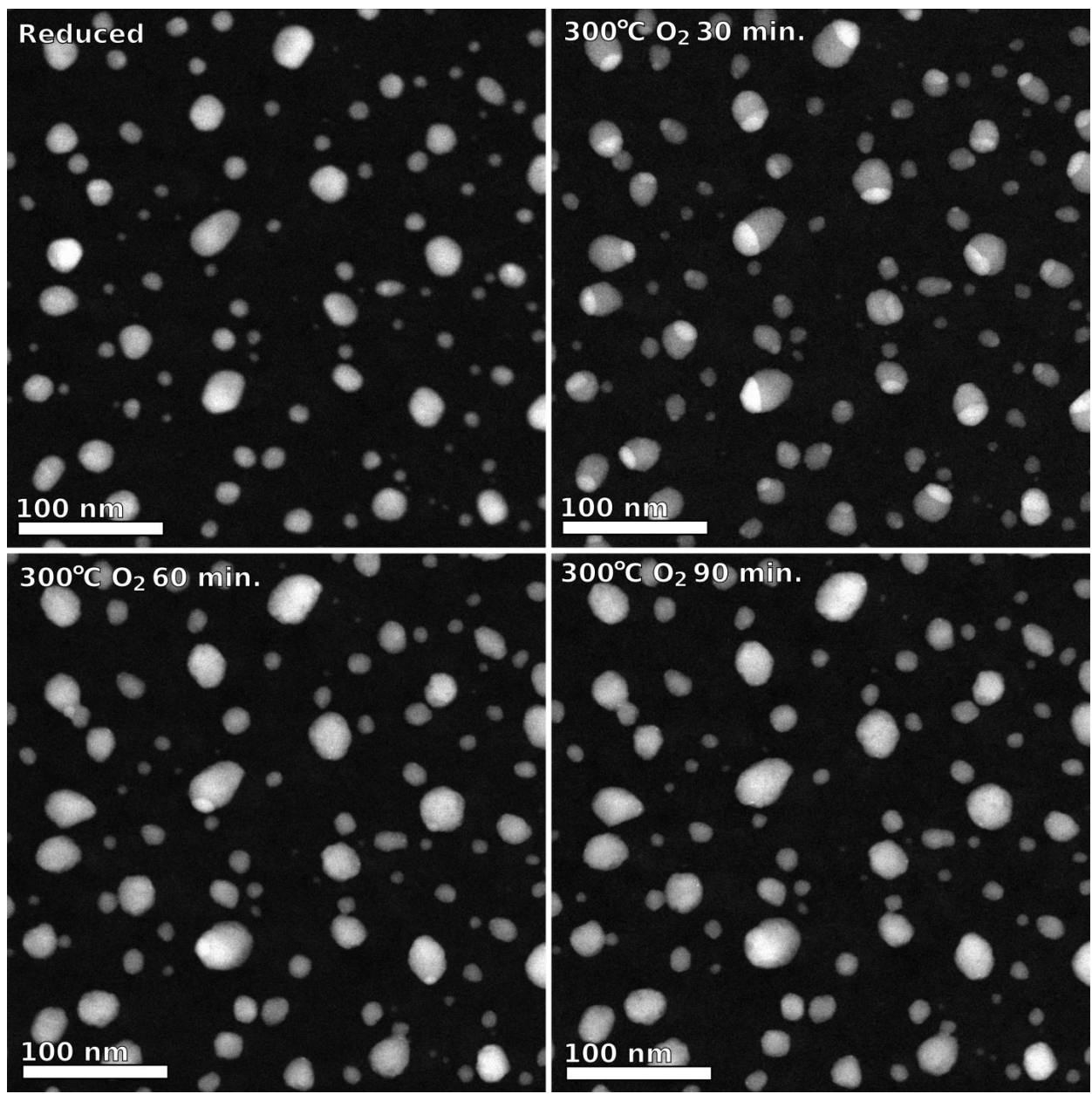

**Figure S1.** *In situ* oxidation of copper nanoparticles on a silicon nitride support at 300°C in 2 Pa oxygen.

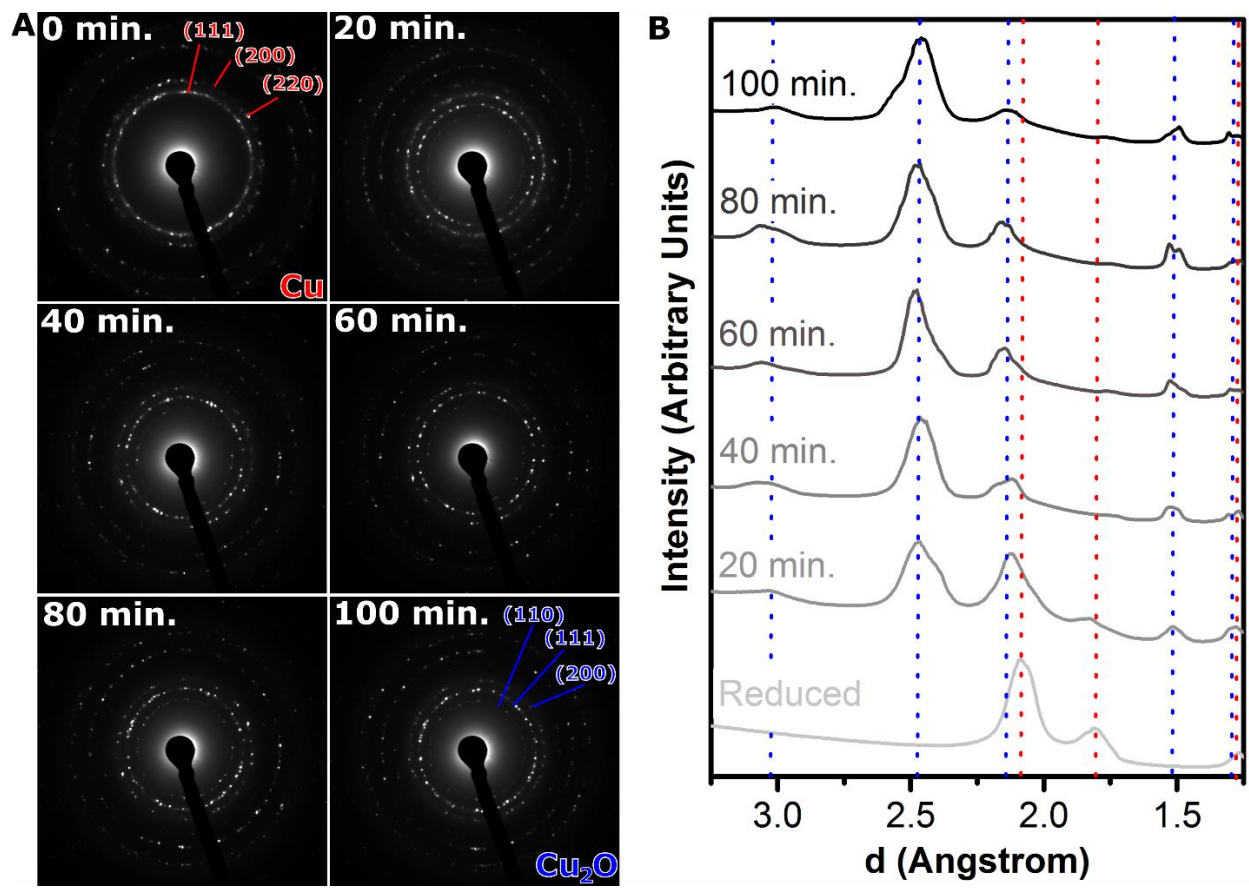

**Figure S2. A)** *In situ* electron diffraction patterns taken over the course of 100 minutes in 2 Pa oxygen with the Cu reflections in red and the $Cu_2O$ reflections in blue. **B)** The radially averaged electron diffraction patterns from **A)** with red and blue lines indicating where diffraction peaks are in Cu and $Cu_2O$ respectively. It should be noted that the electron diffraction was taken from a different area from in **Figure 1**, as the reaction was carried out in ETEM.

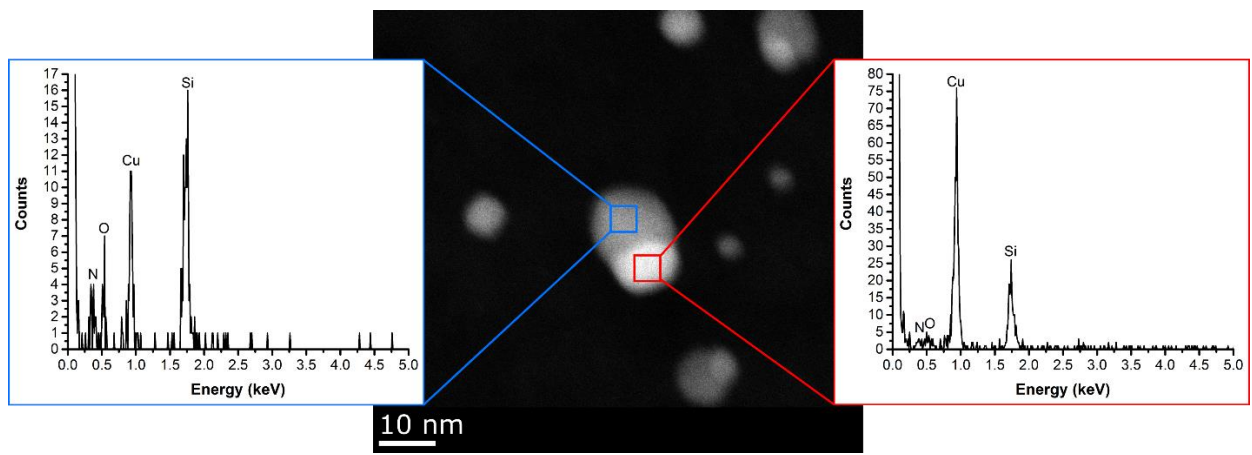

**Figure S3.** Energy dispersive x-ray spectroscopy (EDS) spots from the dark area and bright area showing oxygen present in the darker area. Shorter exposure times were taken in the oxygen region due to beam damage occurring in the copper oxide.

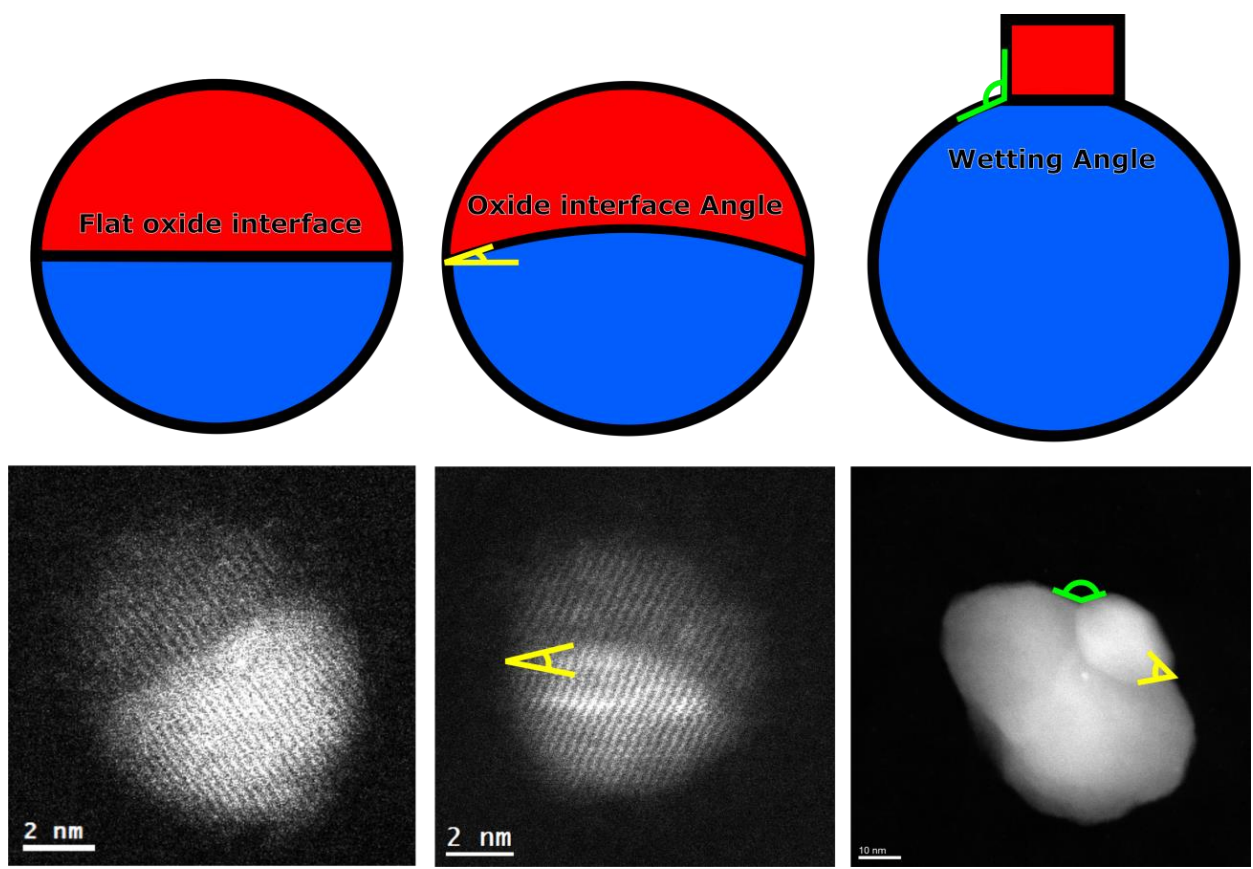

**Figure S4.** A schematic of how the interface angles and wetting angles were defined and measured, and examples of the different cases taken from oxidation on the left and center, and from the reduction on the right.

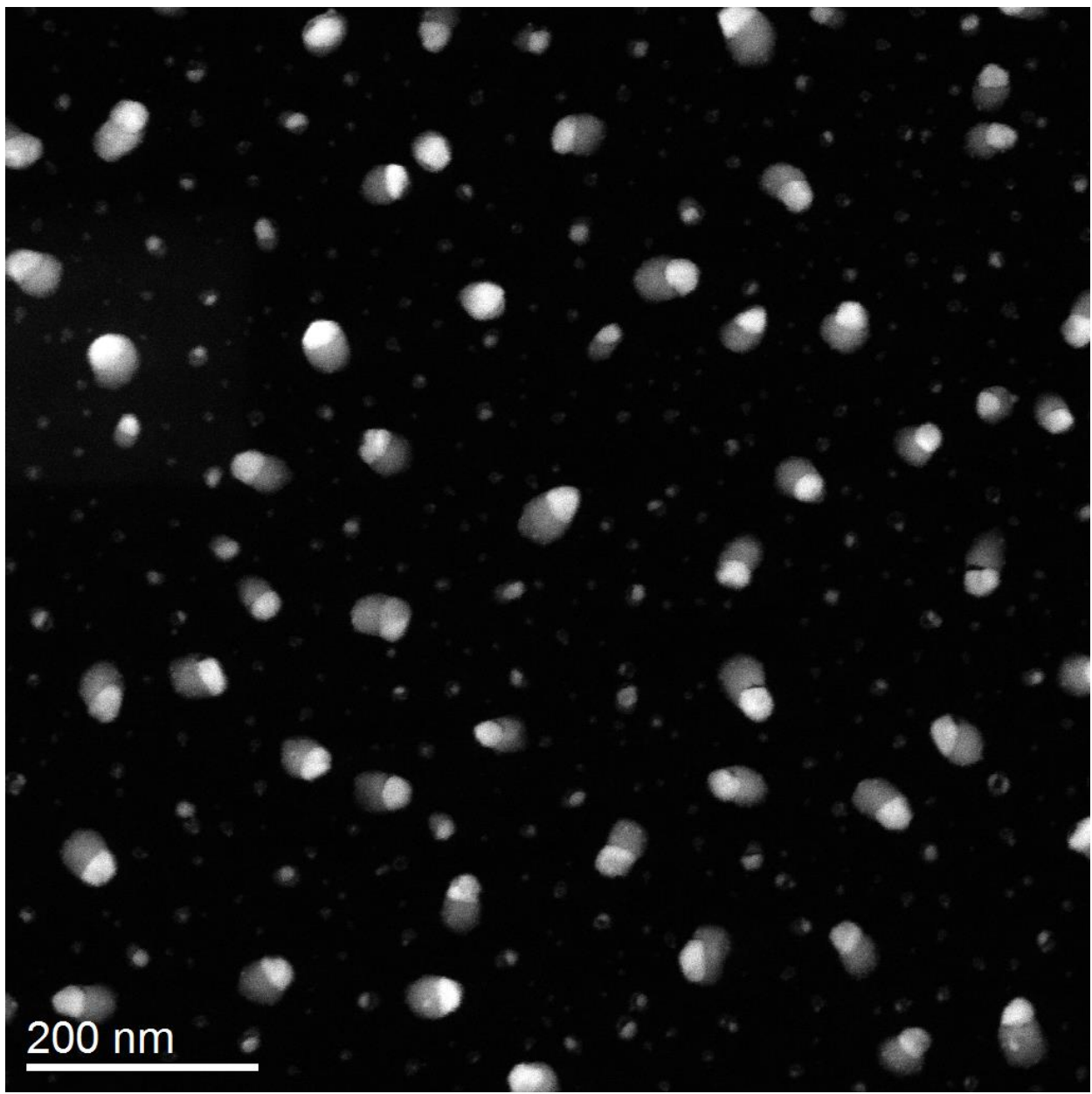

**Figure S5.** *Ex situ* oxidation of copper nanoparticles on a silicon nitride support in a desiccator under ambient conditions after 1 week.

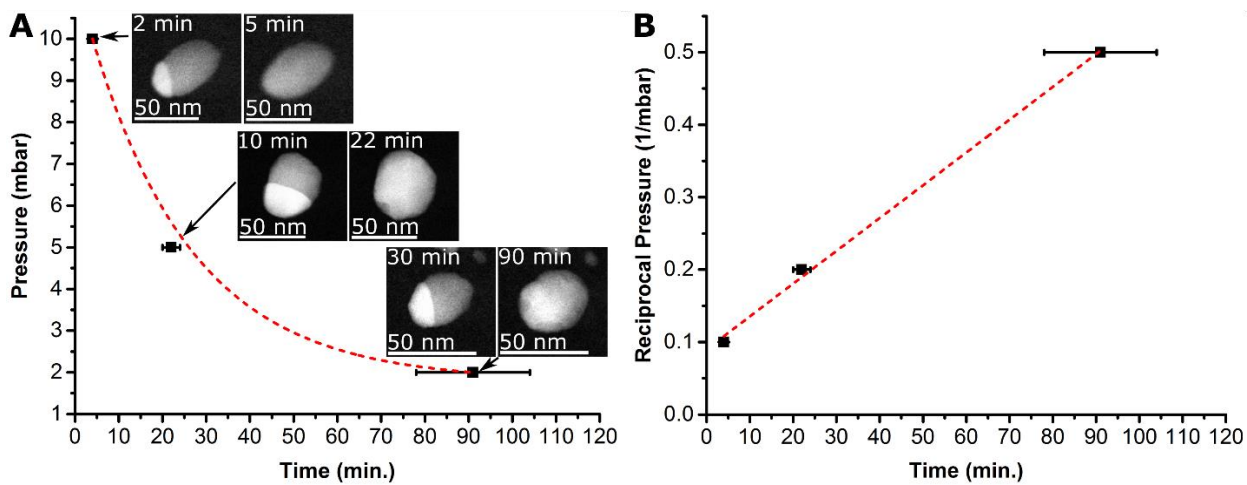

**Figure S6.** Pressure dependence of the initial oxidation stage, with the time of oxidation being shown for ~40 nm nanoparticles at the time when the particle no longer had any of the metallic copper still visible (the right hand image for each point). The red dashed line is a guide added to illustrate the exponential relationship between the points in A), and along the linear relationship vs reciprocal pressure in B).

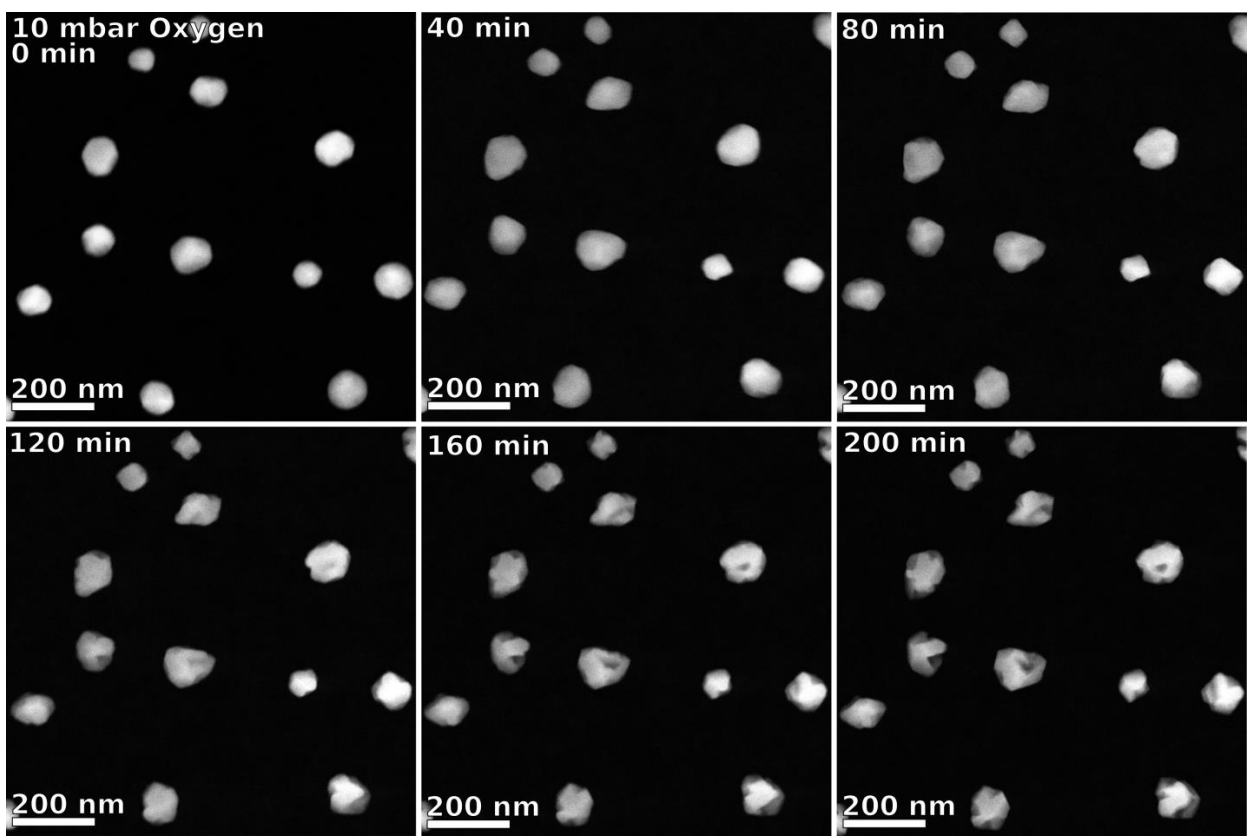

**Figure S7.** *In situ* oxidation of copper nanoparticles at 10 Pa oxygen pressure at 300°C. It should be noted that the initial oxidation occurs within 5 minutes and thus is not seen in this series.

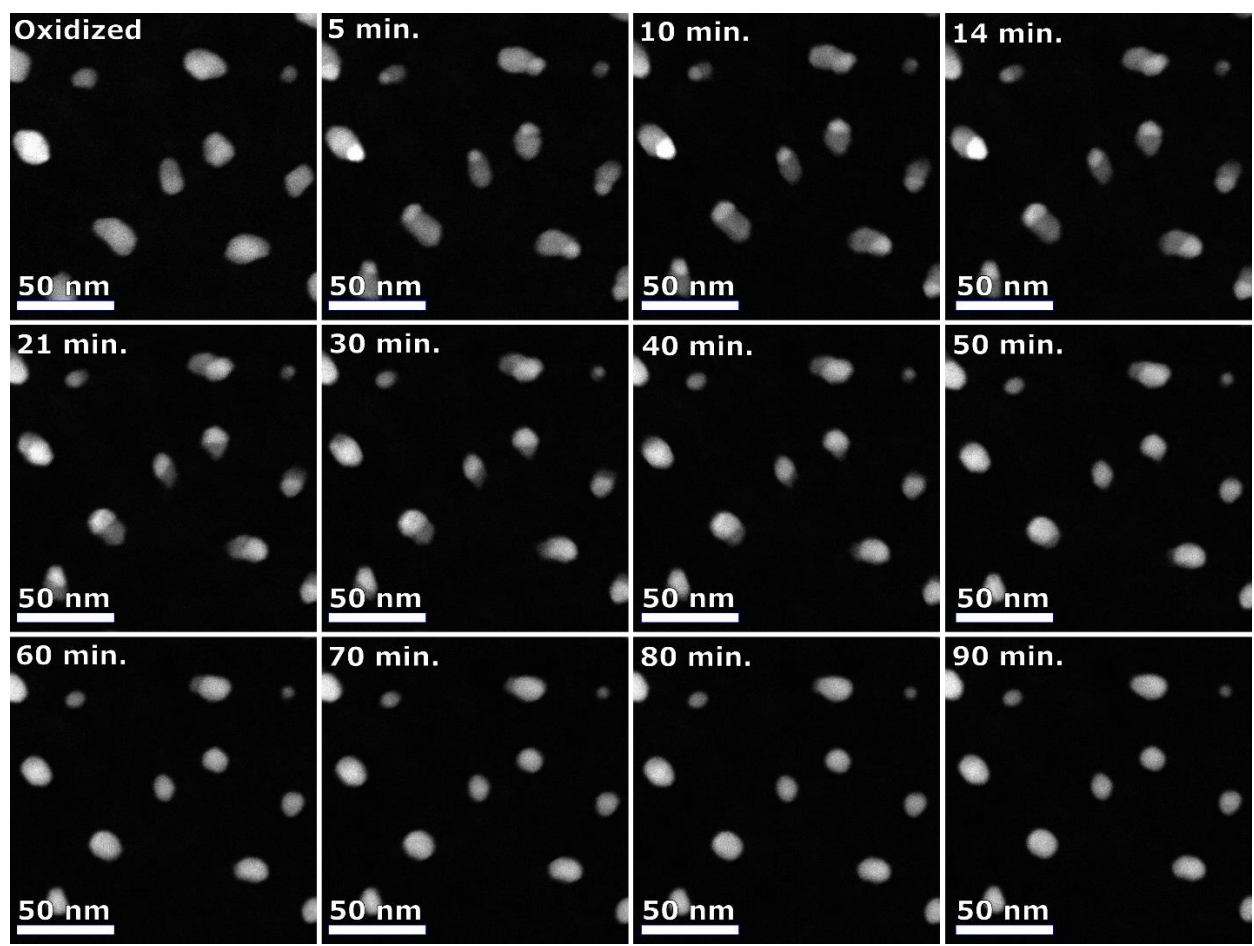

**Figure S8.** The *in situ* reduction of $Cu_2O$ nanoparticles at 400°C in 2 Pa hydrogen on silicon nitride.

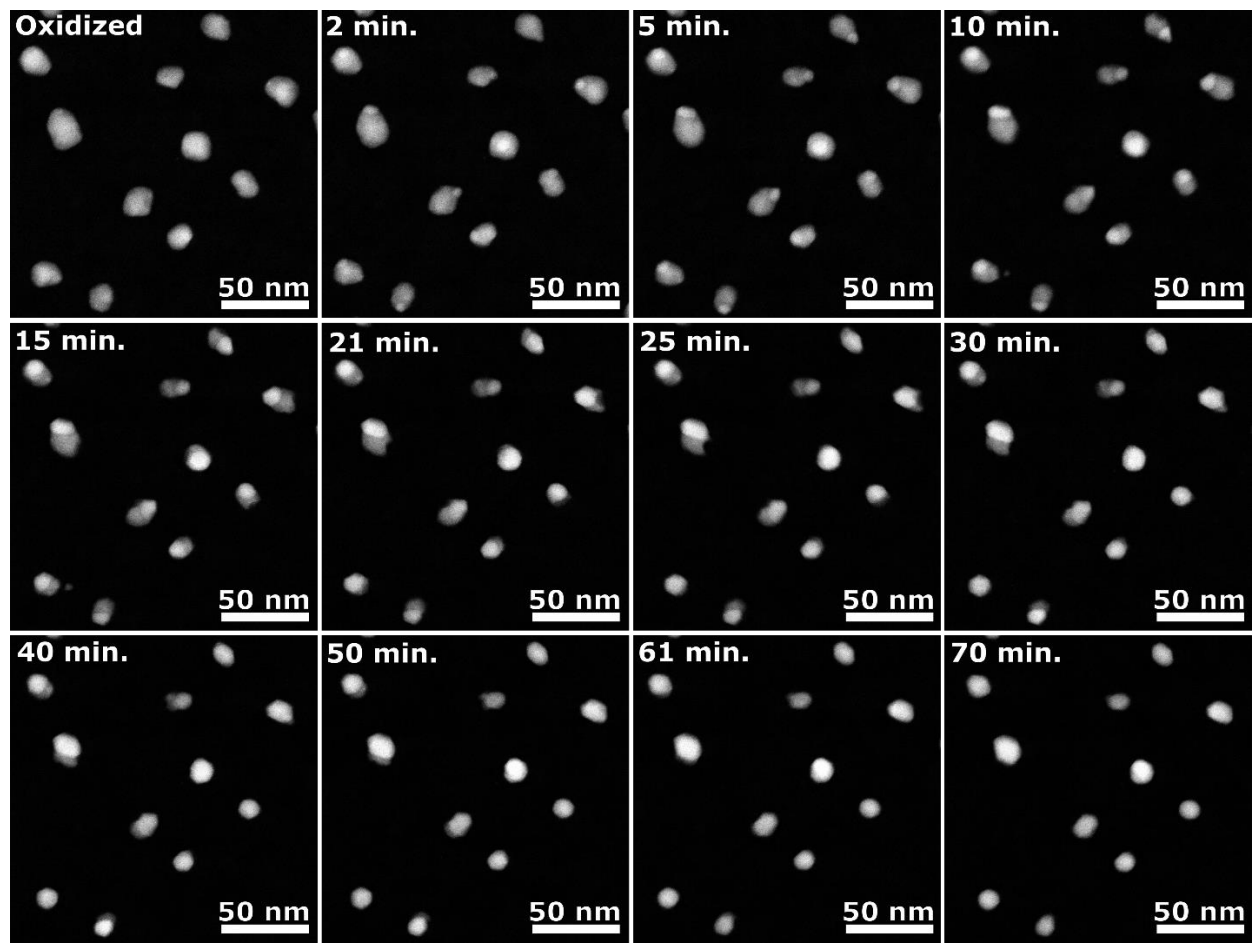

**Figure S9.** The *in situ* reduction of $Cu_2O$ nanoparticles at 500°C in 2 Pa hydrogen on silicon nitride.

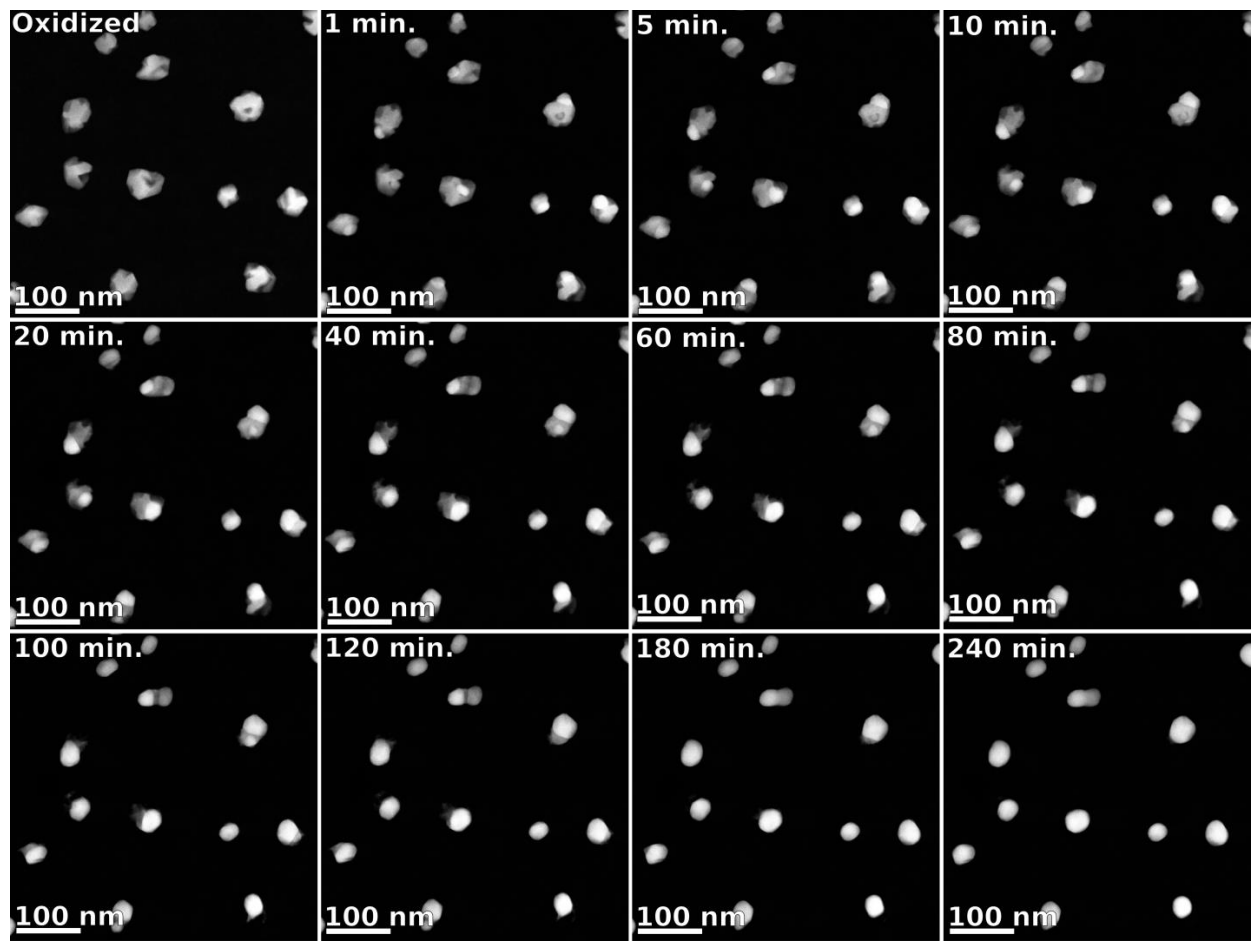

**Figure S10.** The *in situ* reduction at 400°C in 2 Pa hydrogen of the copper oxide particles formed in **Figure S6**.

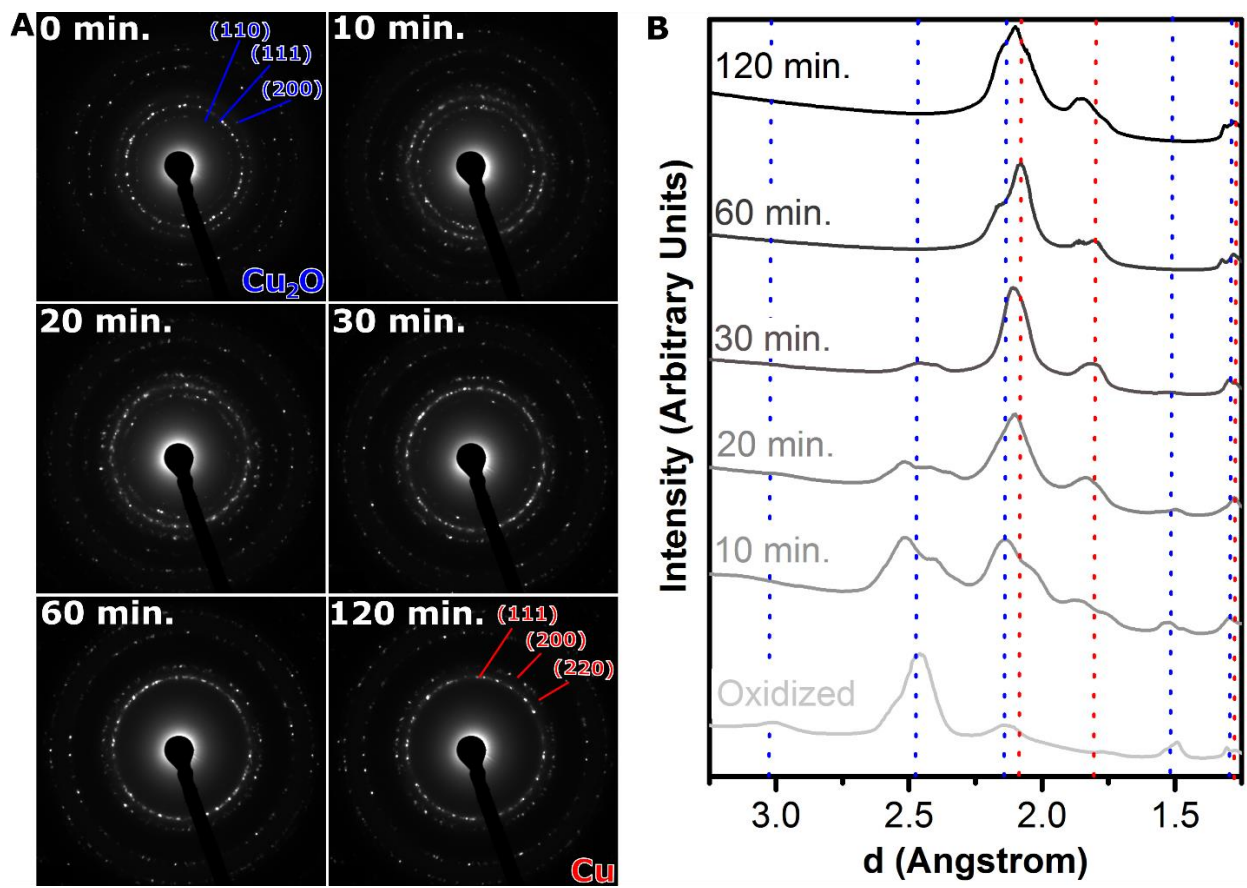

**Figure S11. A)** *In situ* electron diffraction patterns in 2 Pa hydrogen at 400°C and **B)** the radially averaged diffraction patterns, with guides in blue for the *fcc* reflections of $Cu_2O$, and guides in red showing the *fcc* reflections of Cu.

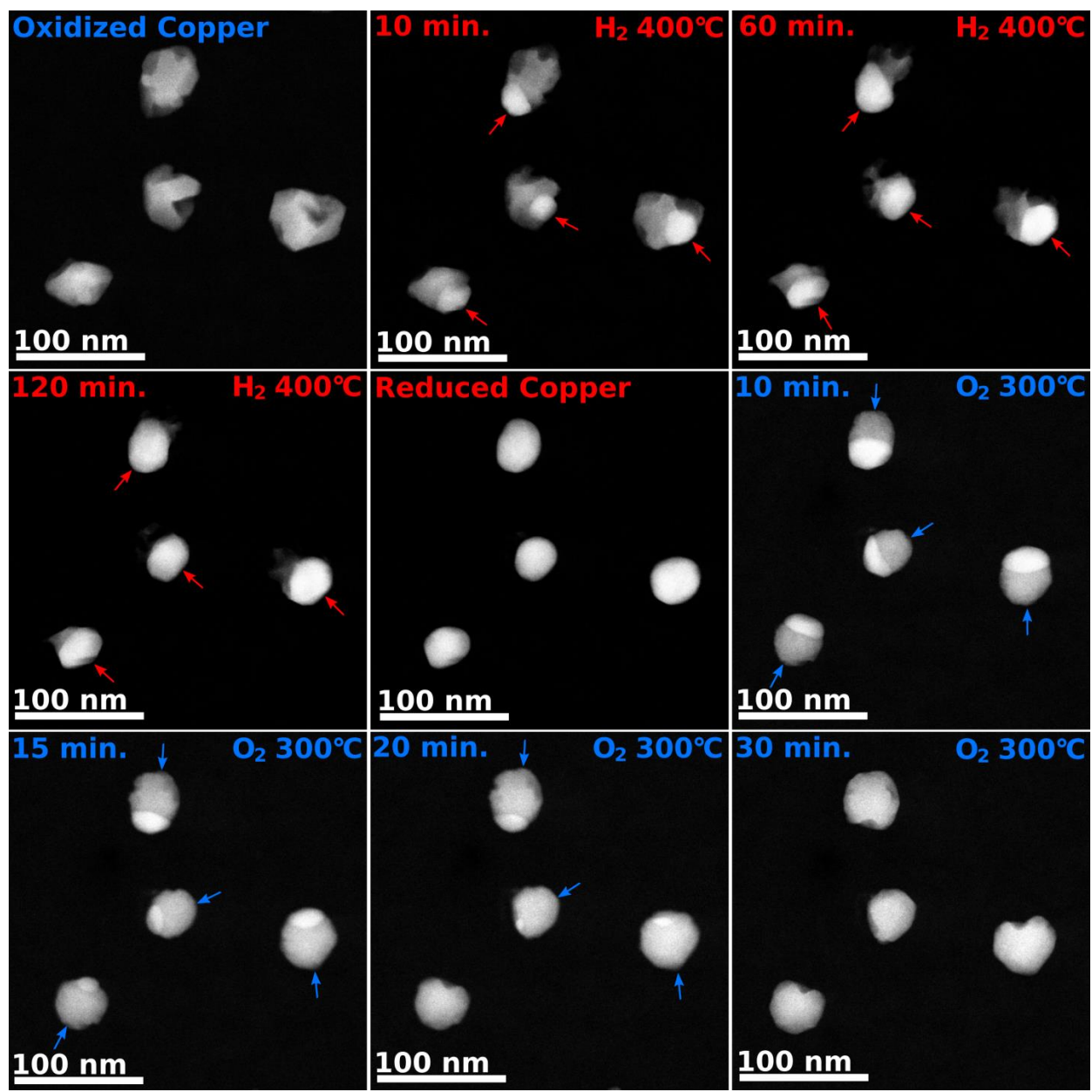

**Figure S12.** A reduction series started with the copper that had been oxidized at 10 Pa oxygen at 300°C for 200 minutes, which is then reduced for 240 minutes in 2 Pa hydrogen at 400°C, and then oxidized in 5 Pa oxygen at 300°C for 30 minutes. The red arrows show the direction of reduction, and the blue arrows are the direction of oxidation.

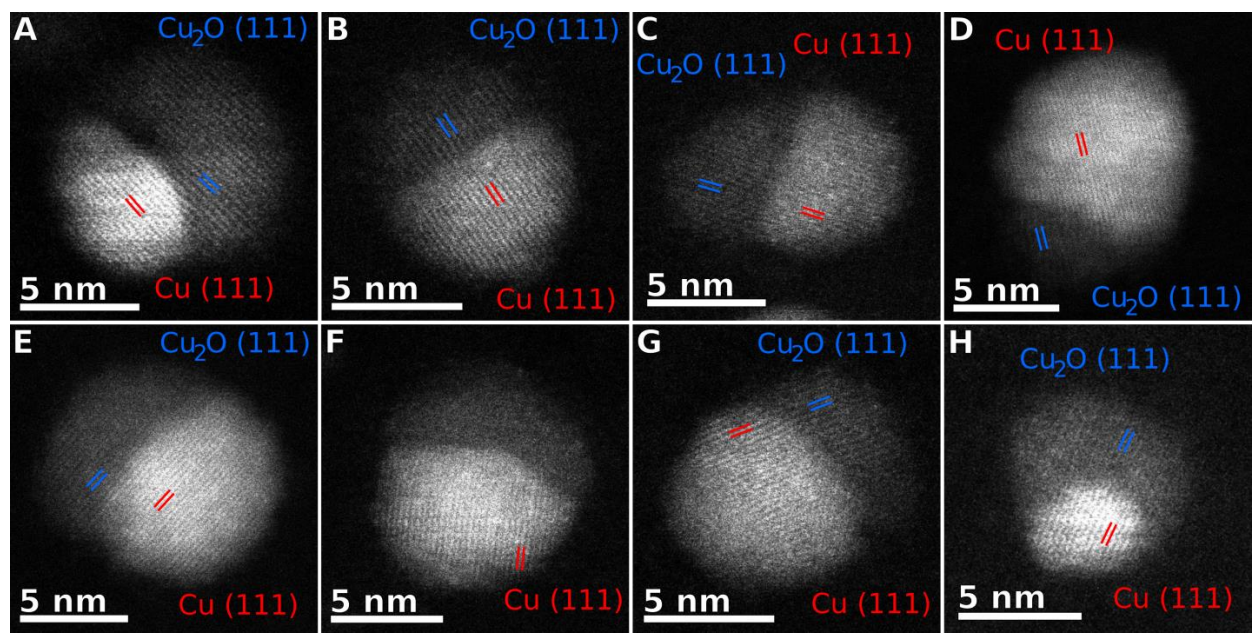

**Figure S13.** HAADF-STEM images of nanoparticles showing the Cu{111}//Cu$_2$O{111} relationship in multiple nanoparticles.

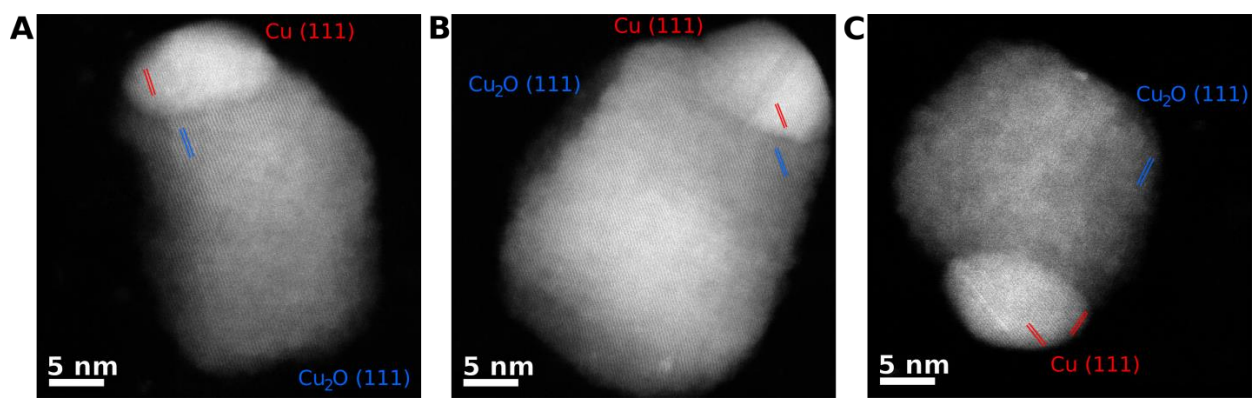

**Figure S14.** Cu/Cu$_2$O nanoparticles formed during the reduction of Cu$_2$O with the (111) Cu$_2$O planes highlighted in blue and the (111) planes of Cu in red.

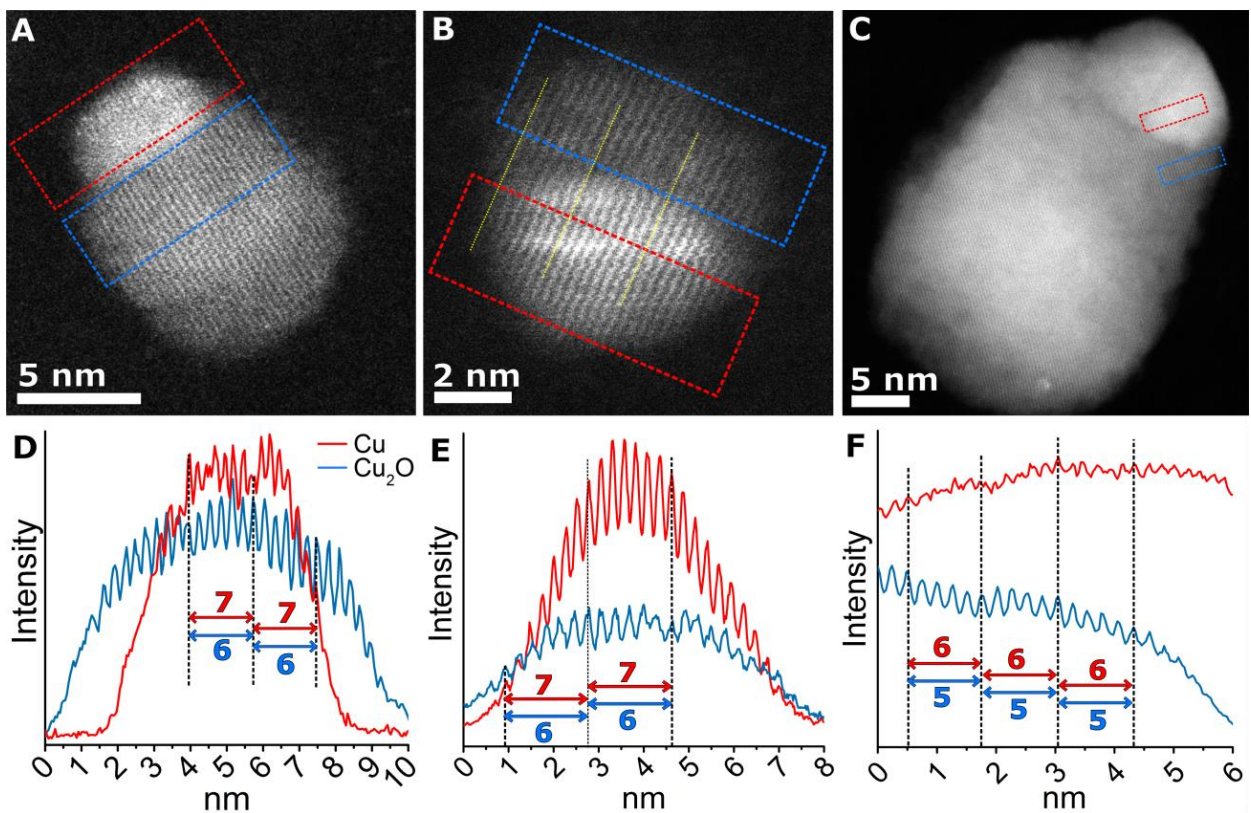

**Figure S15.** Lattice matching across a particle **A)** during oxidation with an interface of ~6 nm, **B)** of 7.4 nm, and **C)** a particle during reduction. The line profiles of **A-C)** are shown respectively in **D-F)**. The red lines come from the red boxes, and the blue lines from the blue boxes. The yellow lines in **B)** show the lattice matching in the image.

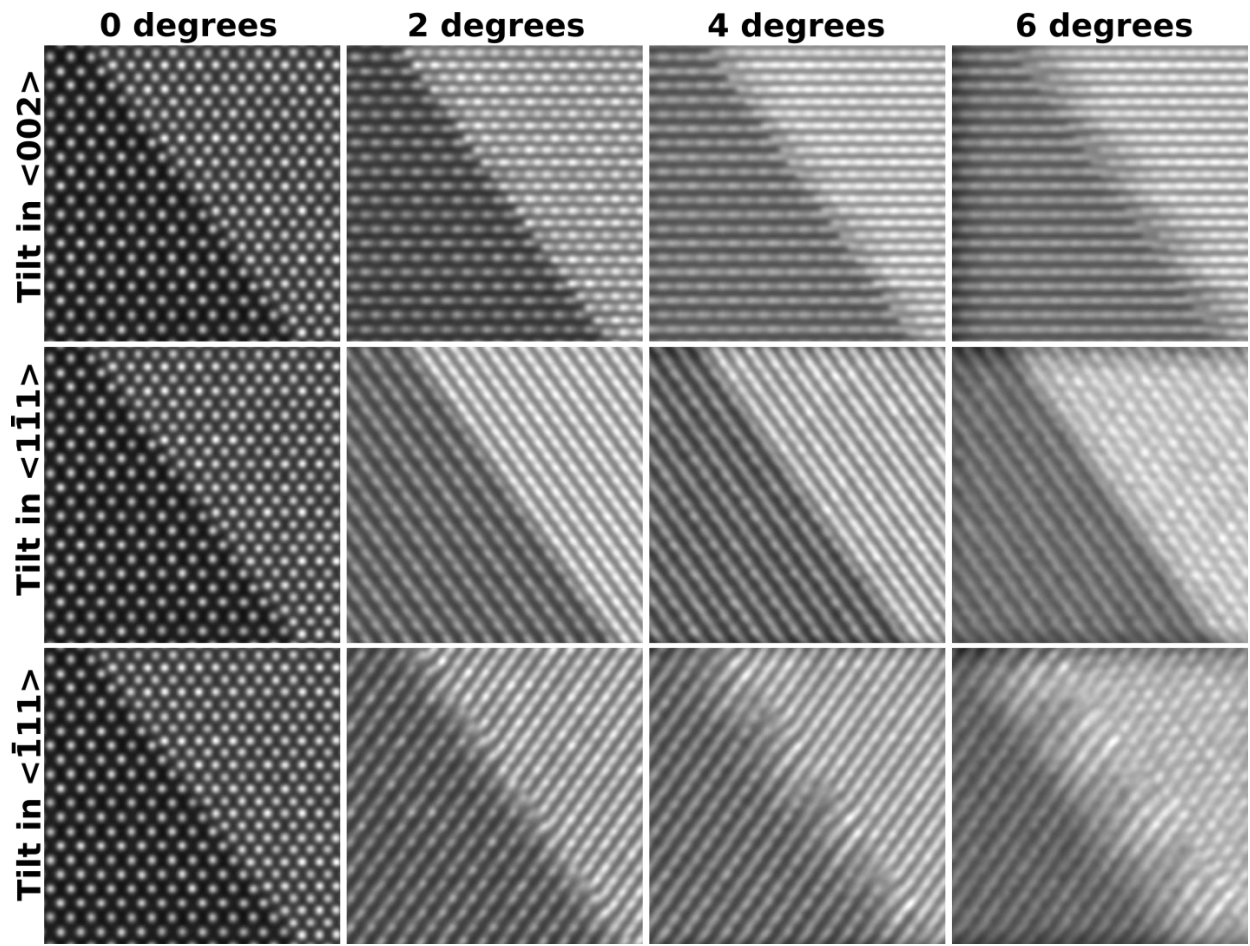

**Figure 16.** QSTEM simulations of the Cu/Cu$_2$O interface rotated along the X direction in the <002> direction, the X and Y direction down the <1-11> direction and the X and –Y direction down the <1-11> direction.

Intensity variations observed between similar atomic columns within the simulation images (**Figure 5 E** and **Figure S16**) are caused by the effects of thermal diffuse scattering. QSTEM's frozen phonon iterations add a small random deviation to each atom position resulting in identical columns having slightly different intensities. It should be noted that tilting the model changes the effective thickness and channeling effects are reduced as atoms no longer form columns. The unusual contrast seen in the high tilt angles in **Figure S16** are likely to be due to limitations of the model where the lattice planes from either side of the interface line up, resulting in disjointed lattice fringes resembling dislocations. The model does not take into account any lattice mismatch relief mechanisms such as strain fields or dislocations formed across the interface.

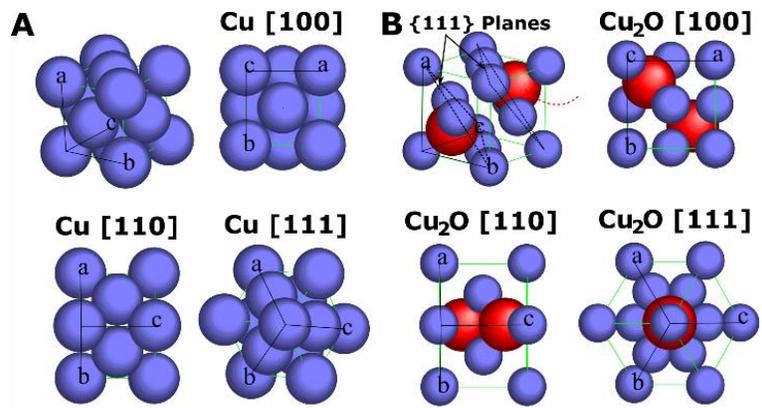

**Figure S17.** Crystal structures of **A)** Cu and **B)** $Cu_2O$ viewed in 3-dimensional space and in the [100], [110], and [111] zone axes.